\documentclass{aip-cp}

\usepackage[numbers]{natbib}
\usepackage{rotating}
\usepackage{graphicx}
\usepackage{epsfig}

\newcommand{\prl}{Phys. Rev. Lett.}
\newcommand{\apj}{Astrophys. J}

\newcommand{\apjl}{Astrophys. J. Lett.}
\newcommand{\apjs}{Astrophys. J. Supp.}

\newcommand{\mnras}{MNRAS}

\newcommand{\prd}{Phys. Rev. D}
\newcommand{\prc}{Phys. Rev. C}
\newcommand{\araa}{An. Rev. Astron. Astrophys.}

\newcommand{\pasa}{Publ. Astron. Soc. Aust.}

\begin{document}

\title{Equation-of-state Constraints and the QCD Phase Transition in the Era of Gravitational-Wave Astronomy}

\author[aff1]{Andreas Bauswein\corref{cor1}}

\author[aff2]{Niels-Uwe Friedrich Bastian}
\eaddress{bastian.niels-uwe@ift.uni.wroc.pl}

\author[aff2,aff3]{David Blaschke}
\eaddress{david.blaschke@ift.uni.wroc.pl}

\author[aff5]{Katerina Chatziioannou}
\eaddress{kchatziioannou@flatironinstitute.org}

\author[aff9]{James Alexander Clark}
\eaddress{james.clark@physics.gatech.edu}

\author[aff2]{Tobias Fischer}
\eaddress{tobias.fischer@ift.uni.wroc.pl}

\author[aff8]{Hans-Thomas Janka}
\eaddress{thj@mpa-garching.mpg.de}

\author[aff6]{Oliver Just}
\eaddress{oliver.just@riken.jp}

\author[aff10]{Micaela Oertel}\eaddress{micaela.oertel@obspm.fr}

\author[aff7]{Nikolaos Stergioulas}
\eaddress{niksterg@auth.gr}


\affil[aff1]{GSI Helmholtzzentrum f\"ur Schwerionenforschung, Planckstra{\ss}e 1, 64291 Darmstadt, Germany}
\affil[aff2]{Institute of Theoretical Physics, University of Wroclaw, Max-Born Pl. 9, 50-204 Wroclaw, Poland}
\affil[aff3]{Bogoliubov Laboratory of Theoretical Physics, JINR Dubna, Joliot-Curie Str. 6, 141980 Dubna, Russia}

\affil[aff5]{Center for Computational Astrophysics, Flatiron Institute, 162 5th Ave, New York, NY 10010}

\affil[aff9]{Center for Relativistic Astrophysics, School of Physics, Georgia Institute of Technology, Atlanta, Georgia 30332, USA}


\affil[aff8]{Max Planck Institute for Astrophysics, Karl-Schwarzschild-Str. 1, 85748 Garching, Germany}

\affil[aff6]{Astrophysical Big Bang Laboratory, RIKEN Cluster for Pioneering Research, 2-1 Hirosawa, Wako, Saitama 351-0198, Japan}

\affil[aff10]{LUTH, CNRS, Observatoire de Paris, PSL Research University, Universit{\'e} Paris Diderot, Sorbonne Paris Cit{\'e}, 5 place Jules Janssen, 92195 Meudon, France} 

\affil[aff7]{Department of Physics, Aristotle University of Thessaloniki, 54124 Thessaloniki, Greece}

\corresp[cor1]{Corresponding author: a.bauswein@gsi.de}

\maketitle

\begin{abstract}
We describe a multi-messenger interpretation of GW170817, which yields a robust lower limit on NS radii. This excludes NSs with radii smaller than about 10.7~km and thus rules out very soft nuclear matter. We stress the potential of this type of constraints when future detections become available. For instance, a very similar argumentation may yield an upper bound on the maximum mass of nonrotating NSs. We also discuss simulations of NS mergers, which undergo a first-order phase transition to quark matter. We point out a different dynamical behavior. Considering the gravitational-wave signal, we identify an unambiguous signature of the QCD phase transition in NS mergers. We show that the occurrence of quark matter through a strong first-order phase transition during merging leads to a characteristic shift of the dominant postmerger frequency. The frequency shift is indicative for a phase transition if it is compared to the postmerger frequency which is expected for purely hadronic EoS models. A very strong deviation of several 100~Hz is observed for hybrid EoSs in an otherwise tight relation between the tidal deformability and the postmerger frequency. In future events the tidal deformability will be inferred with sufficient precision from the premerger phase, while the dominant postmerger frequency can be obtained when current detectors reach a higher sensitivity in the high-frequency range within the next years. Finally, we address the potential impact of a first-order phase transition on the electromagnetic counterpart of NS mergers. Our simulations suggest that there would be no significant qualitative differences between a system undergoing a phase transition to quark matter and purely hadronic mergers. The quantitative differences are within the spread which is found between different hadronic EoS models. This implies on the one hand that GW170817 is compatible with a possible transition to quark matter. On the other hand these considerations show that it may not be easy to identify quantitative differences between purely hadronic mergers and events in which quark matter occurs considering solely their electromagnetic counterpart or their nucleosynthesis products.

\end{abstract}

\section{INTRODUCTION}
The observation of gravitational waves has nearly become routine. By now 10 binary black hole mergers have been detected and one system with two NSs has been found~\cite{Abbott2017,Collaboration2018}. The latter dubbed GW170817 referring to its detection date on August 17, 2017 is the by far most interesting measurement for understanding the physics of high-density matter and for nuclear astrophysics especially because emission from the electromagnetic counterpart was observed~\cite{Abbott2017b}. The total mass of $M_\mathrm{tot}=M_1+M_2\approx2.7~M_\odot$, which was inferred from the GW signal, is compatible with masses measured in binary NS systems containing pulsars \cite{Oezel2016}. The binary mass ratio $q=M_1/M_2$ was constrained to be between 0.7 and 1. The extraction of finite-size effects during the GW inspiral phase put a constraint on the tidal deformabiltiy~\cite{Abbott2017,Abbott2019,PhysRevLett.121.161101,Chatziioannou2018,De2018,Carney2018}. The strong correlation between the tidal deformability and NS radii implies that NS radii of typical masses can not be larger than about 13.5~km (slightly depending on assumptions made during the analysis). The measurement thus implies that nuclear matter cannot be very stiff, which is inline with current knowledge from nuclear physics (see e.g.~\cite{Lim2018,Tews2019}).

The properties of the electromagnetic counterpart of GW170817 in the optical and infrared (e.g.~\cite{Villar2017,Cote2018}) is compatible with unbound matter which is heated by radioactive decays of products of the rapid neutron-capture process~\cite{Metzger2010}. Although many details still have to be clarified like the exact amount of the ejecta or their composition, the data provides very strong evidence that NS mergers play an important role for the enrichment of the Universe by heavy elements~\cite{Lattimer1977}.

In this contribution we address two aspects of NS mergers. First we argue that a multi-messenger interpretation of GW170817 leads to an additional constraint on the properties of nuclear matter. Using a minimum set of assumptions we show that NS radii of typical masses should be larger than about 10.7~km meaning that high-density matter cannot be very soft. We point out that this new method for radius and EoS constraints bears a lot of potential for the future when new measurements become available. Second, we describe merger simulations with EoS models which feature a first-order phase transition to quark matter. We identify an unambiguous signature of a strong first-order phase transition as it is expected to occur, when quarks are becoming deconfined at some higher density, which may or may not take place in NSs. The scenario requires a sufficiently precise detection of the tidal deformability and the measurement of the dominant postmerger GW frequency. Hence, the identification of such a characteristic signature may be achieved in a few years from now when instruments with higher sensitivity, in particular in the kHz range, become operational. Our findings demonstrate the complementary information buried in the GW signals of the inspiral and the postmerger phase. Moreover, we discuss based on our simulations the possible impact of a phase transition to quark matter on the electromagnetic counterpart of NS mergers.

%

\section{Radius constraint from GW170817}
GW170817 was accompanied by electromagnetic emission from gamma rays to radio~\cite{Abbott2017b}. The evolution of the light curve in the infrared and optical permits the inference of ejecta properties of the merger as the radiation at these wavelength is likely powered by radioactive decays in the unbound material. The inferred numbers are somewhat model-dependent, but in generally good agreement with expectations from numerical simulations of the merger process and the long-term postmerger evolution. The ejecta masses derived by different groups (see e.g.~\cite{Cote2018} for a compilation) are relatively high, but compatible with theoretical models, e.g.~\cite{Wu2016}.

In~\cite{Bauswein2017} we argued that this observation provides evidence that the merger did not directly form a black hole. This argument is based on simulation results showing that prompt black-hole formation leads to a reduction of the ejecta mass, e.g.~\cite{Bauswein2013a,Hotokezaka2013}. See also discussion in~\cite{Ai2018} and references therein. We further argued that this interpretation if correct implies a lower bound on NS radii. The reason is that the threshold binary mass $M_\mathrm{thres}$ for prompt black-hole formation in NS mergers depends in a particular way on the EoS. Characterizing the EoS through stellar properties of nonrotating NSs, simulations show that $M_\mathrm{thres}$ is to good approximation given by~\cite{Bauswein2013}
\begin{equation}
M_\mathrm{thres}= \left( -3.606\frac{G\, M_\mathrm{max}}{c^2\,R_{1.6}}+2.38\right)\,M_\mathrm{max}
\end{equation}
with $R_{1.6}$ being the radius of a nonrotating NS with 1.6~$M_\odot$ and $M_\mathrm{max}$ being the maximum mass of nonrotating NSs. If GW170817 did not result in the direct formation of a black hole, its measured total binary mass should be smaller than $M_\mathrm{thres}$ because only binaries with masses above the threshold mass collapse promptly. Hence, we obtain
\begin{equation}\label{eq:comp}
M_\mathrm{tot}^\mathrm{GW170817}=2.74~M_\odot < M_\mathrm{thres} = \left( -3.606\frac{G\, M_\mathrm{max}}{c^2\,R_{1.6}}+2.38\right)\,M_\mathrm{max}.
\end{equation}
Both, $R_{1.6}$ and $M_\mathrm{max}$, are not precisely known. But, they are not fully uncorrelated. An EoS yielding a given $R_{1.6}$, cannot have an arbitrarily high maximum mass because causality requires that the speed of sound is smaller than the speed of light and thus limits the stiffness of the EoS. This means that $M_\mathrm{max}$ is constrained by a given $R_{1.6}$. This limit can be found empirically by considering a large number of different EoSs, which are modified at higher densities such that the EoS becomes maximally stiff. The constraint is found to be given by
\begin{equation}\label{eq:limit}
M_\mathrm{max}<\frac{1}{3.1}\frac{c^2}{G}R_{1.6}.
\end{equation}
Inserting the limit~Equation~(\ref{eq:limit}) in Equation~(\ref{eq:comp}) immediately yields
\begin{equation}
R_{1.6}>2.55\frac{G}{c^2}M_\mathrm{tot}^\mathrm{GW170817}.
\end{equation}
Refining the argument by assuming a remnant lifetime of at least 10~ms results in $R_{1.6}>10.7$~km. For a discussion of more details and the error budget see~\cite{Bauswein2017,Bauswein2019a}.

This conclusion implies that nuclear matter cannot be very soft because otherwise the remnant of GW170817 collapsed promptly into a black hole which is arguable disfavored by the electromagnetic display. Note that our limit represents a rather robust constraint because we made conservative assumptions throughout and for instance imposed that the stiffness of the EoS is only limited by causality. Importantly, our constraint is complementary to the limit given by the tidal deformability, which yields an upper limit on NS radii~\cite{Abbott2017,Abbott2019,PhysRevLett.121.161101,Chatziioannou2018,De2018,Carney2018}. 

\begin{figure}[h]
  \includegraphics[width=250pt]{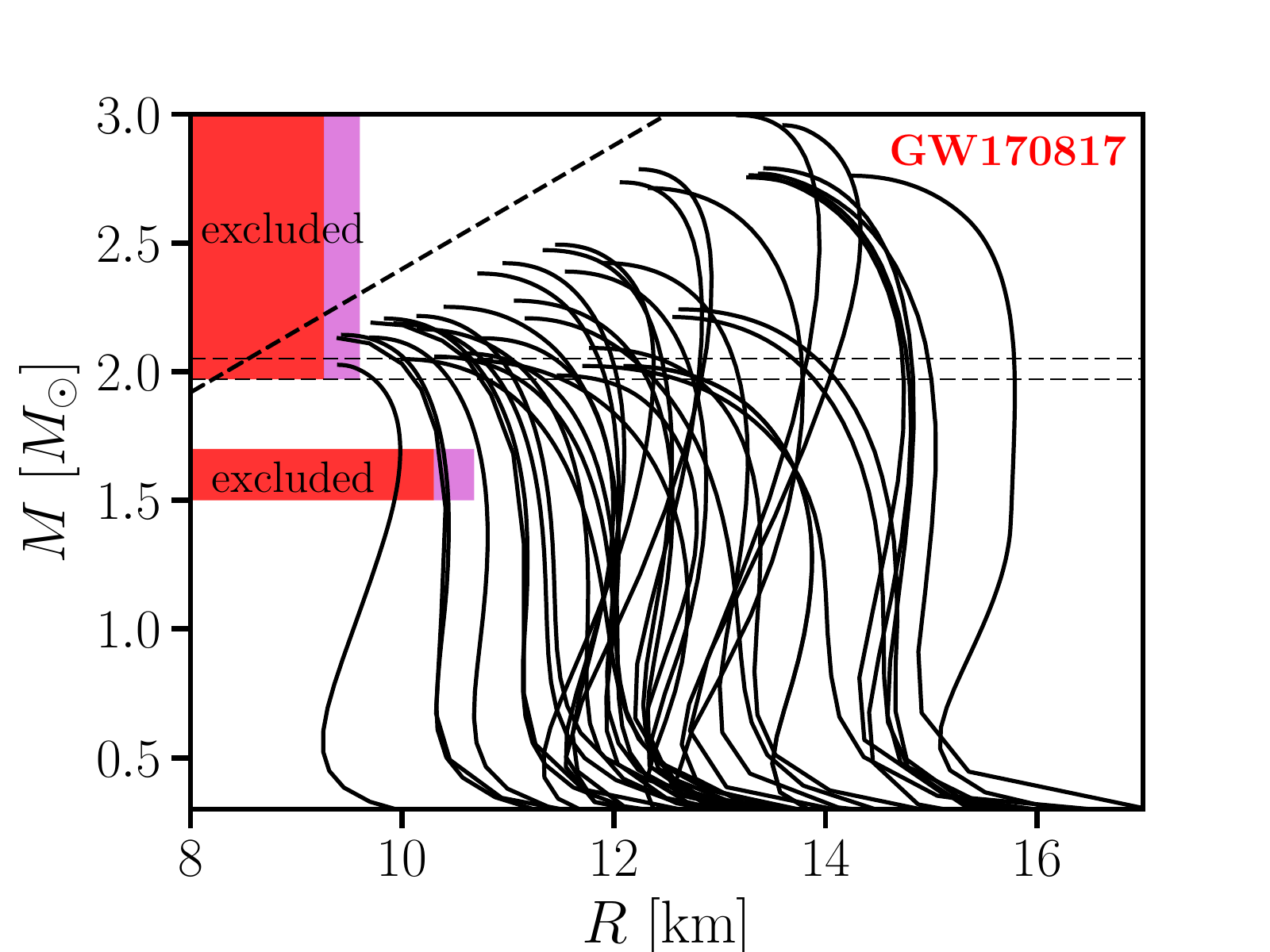}
  \caption{Mass-radius relations of nonrotating NSs for different EoS models overplotted with our constraints on NS radii. Very soft EoSs are excluded. The horizontal line indicates the lower limit on the maximum mass of nonrotating NSs set by~\cite{Antoniadis2013}. The diagonal dashed line shows the causality limit. Figure taken from~\cite{Bauswein2017}. \label{fig:eoscon}}
\end{figure} 

Our constraint is visualized in Figure~\ref{fig:eoscon}, which shows different EoSs available in the literature in comparison to our limit. In addition, Figure~\ref{fig:eoscon} indicates a bound on $R_\mathrm{max}$, the radius of the maximum-mass configuration with $M=M_\mathrm{max}$. A very similar derivation as above leads to $R_\mathrm{max}>9.6$~km.
 
Two comments are in order. First, our constraints are not very restrictive as only the softest EoSs are ruled out. However, future events with a similar electromagnetic signature but a higher total binary mass will lead to stronger lower limits. As such type of radius constraints require only the measurement of the total binary mass, they can be obtained from any new event even with low signal-to-noise ratio provided the electromagnetic counterpart is identified. 

Also note that a very similar line of arguments can be employed in future events to obtain an upper bound on NS radii and on the maximum mass $M_\mathrm{max}$ if the electromagnetic counterpart provides evidence for a prompt collapse of the merger remnant. Hence, this new method of EoS constraints from a multi-messenger interpretation of NS mergers bears significant potential for the future, see e.g. Figures~3 and~4 in~\cite{Bauswein2017} for hypothetical cases.
\begin{figure}[h]
  \includegraphics[width=250pt]{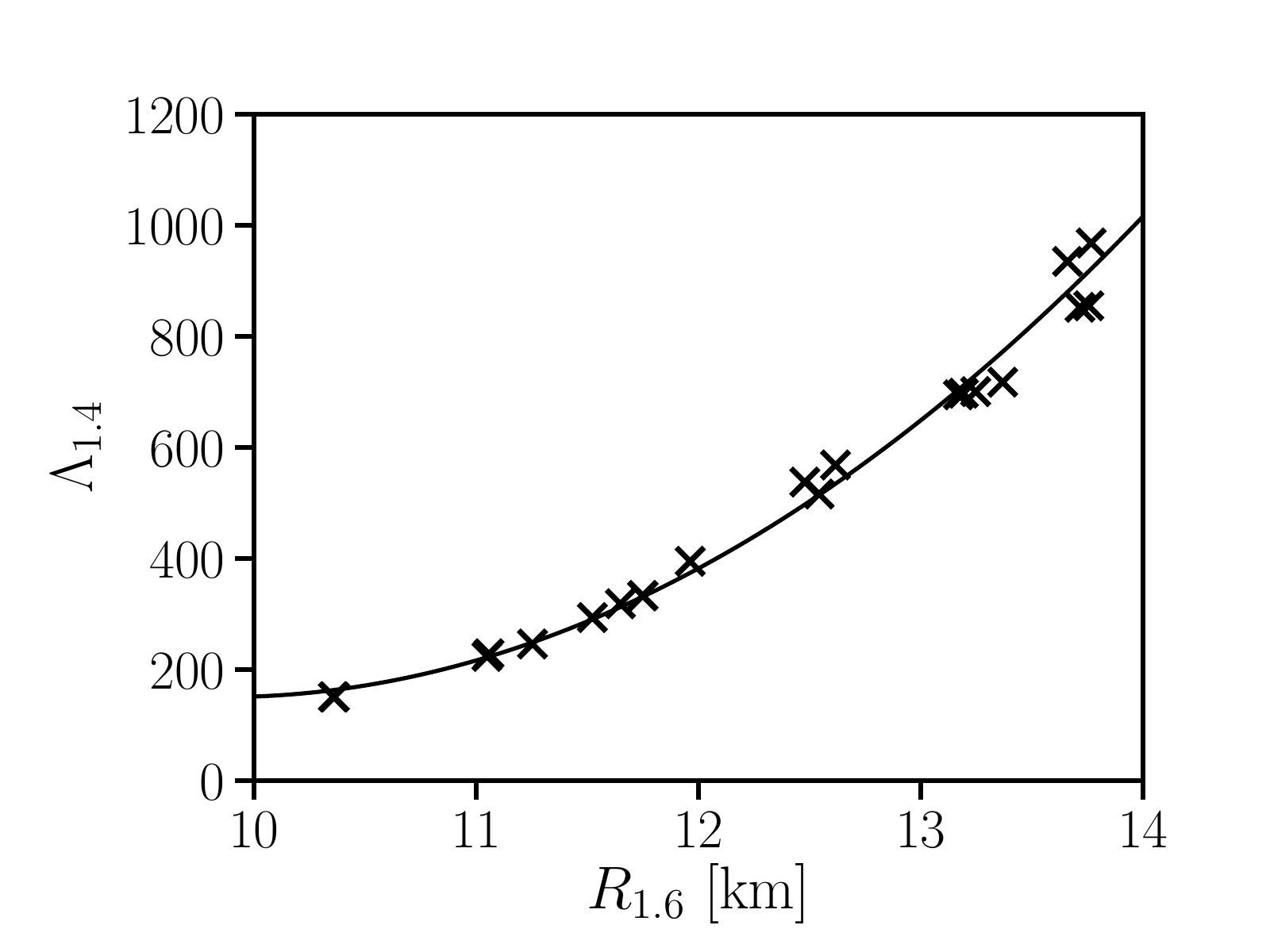}
  \caption{Relation between the radii of nonrotating NSs with 1.6~$M_\odot$ and the tidal deformability of nonrotating NSs with a mass of 1.4~$M_\odot$ for different EoSs. The solid curve provides a second-order polynomial least-square fit describing the tight relation between both quantities.}
\label{fig:lamrad}
\end{figure}

Second, it is very simple to convert our current radius constraint to a limit on the tidal deformability $\Lambda=\frac{2}{3}k_2\left( \frac{c^2 R}{G M} \right)^5$ with the tidal Love number $k_2$~\cite{Hinderer2008}. We compute radii of NSs with 1.6~$M_\odot$ and the tidal deformability $\Lambda_{1.4}$ for NSs with a mass of 1.4~$M_\odot$, which is roughly the mass of the progenitor stars in GW170817. Doing this for a large representative sample of EoSs (see~\cite{Bauswein2019a} or Figure~\ref{fig:lamrad}), one finds a tight relation between these two quantities. This is not surprising considering the strong impact of NS radii on $\Lambda$ and that the radius for a given EoS does not change very strongly for NS masses between 1.4~$M_\odot$ and 1.6~$M_\odot$. Using this relation we conclude that $\Lambda_{1.4}$ should be larger than about 200.

This limit of $\Lambda_{1.4}>200$ can be compared to other studies in the literature which follow very similar arguments. Ref.~\cite{Radice2018b} reported a lower limit of $\Lambda> 300$, which seems stronger than our limit. However, the discrepancy stems from considering only four EoS models in~\cite{Radice2018b}. The study thus misses possible EoS models with smaller tidal deformability which are still compatible with the observational data. In our approach through an empirical relation we ensure a complete coverage of all possible EoSs. We thus stress that the current data implies a lower limit of only $\Lambda_{1.4}>200$, and models with larger tidal deformability are still compatible with the data of GW170817 (see also~\cite{Tews2019,Koeppel2019,Kiuchi2019,Bauswein2019a}). Obviously, new detections with higher total binary masses may strengthen the constraint as discussed above.

\section{QCD phase transition in NS mergers}
\subsection{Equations of state models}

In this section we describe simulations of compact star mergers. We investigate the differences between models which undergo a strong first-order phase transition to quark matter and models that describe purely hadronic matter and do not feature a strong phase transition. The main driver for this study is the question whether a strong first-order phase transition leaves a detectable and unambiguous signature in one of the observables of NS mergers, e.g. the gravitational-wave emission or the mass ejection that generates an electromagnetic counterpart. For other studies discussing phase transitions in the context of compact object mergers or the two-family scenario see e.g~\cite{Csaki2018,Paschalidis2018,Most2019,Han2018,Christian2018,Sieniawska2018,Burgio2018a,Drago2018,Dexheimer2018}.
\begin{figure}[h]
  \includegraphics[width=250pt]{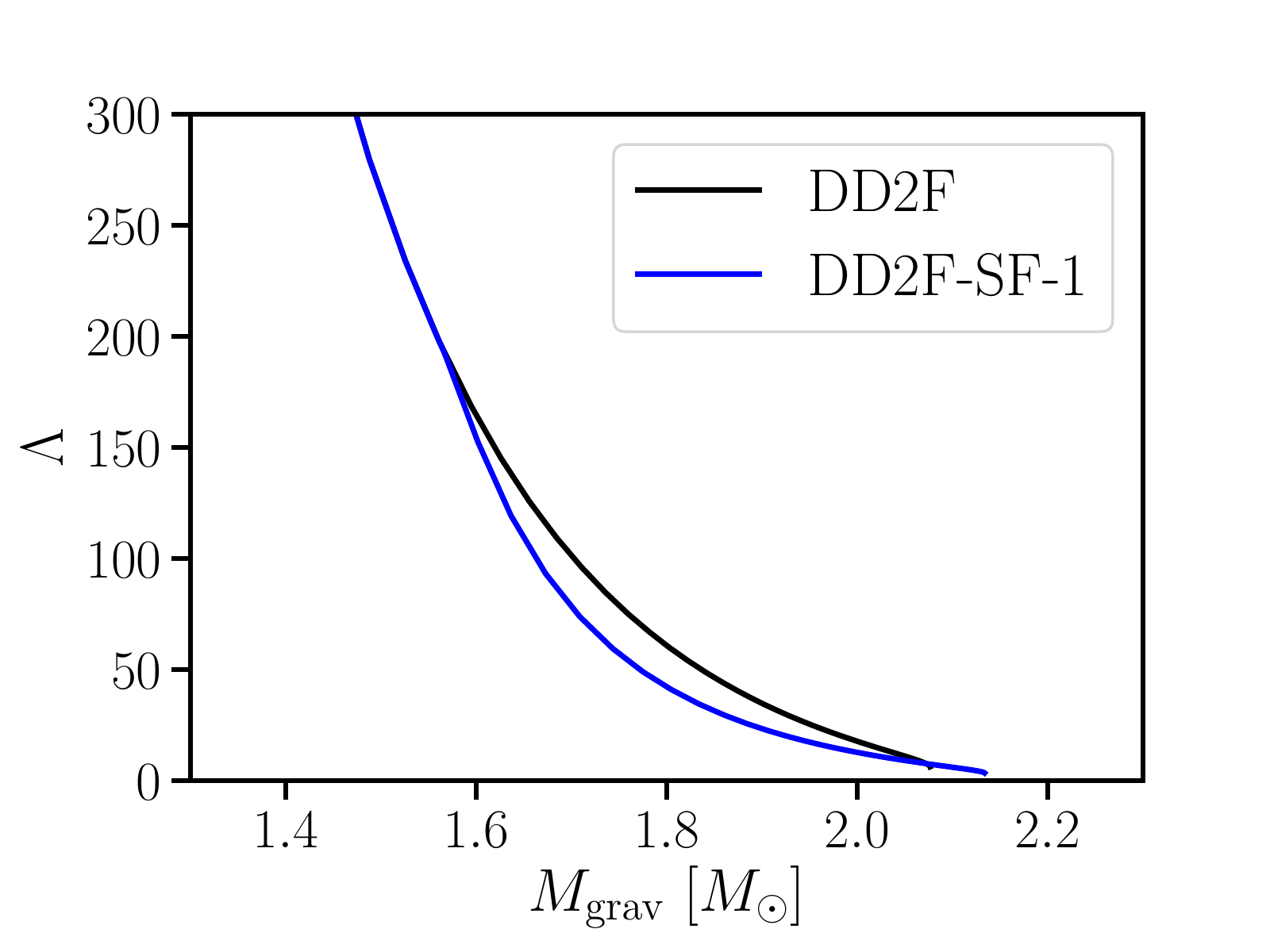}
  \caption{\label{fig:lamm}Tidal deformability as function of mass for a purely nucleonic EoS (black curve) and a model with a first-order phase transition to quark matter (blue curve). Compare also to the mass-radius relations in Figure~\ref{fig:rm}.}
\end{figure} 

We note already here that in principle a first-order phase transition may be directly detectable by considering the tidal deformability of systems with different masses. In Figure~\ref{fig:lamm} the tidal deformability as function of mass exhibits a kink when the phase transition occurs at a mass of about 1.6~$M_\odot$ (more information on the models are given below). However, as the transition takes place at a relatively high mass, it is unclear whether binary systems with such high binary masses are very abundant and whether the tidal deformability can be measured with sufficient precision to actually resolve the kink. Also, the tidal deformability strongly decreases with mass. A small tidal deformability implies less pronounced finite-size effects during the insprial, which makes it more difficult to infer them from the GW signal.

\begin{figure}[h]
  \includegraphics[width=250pt]{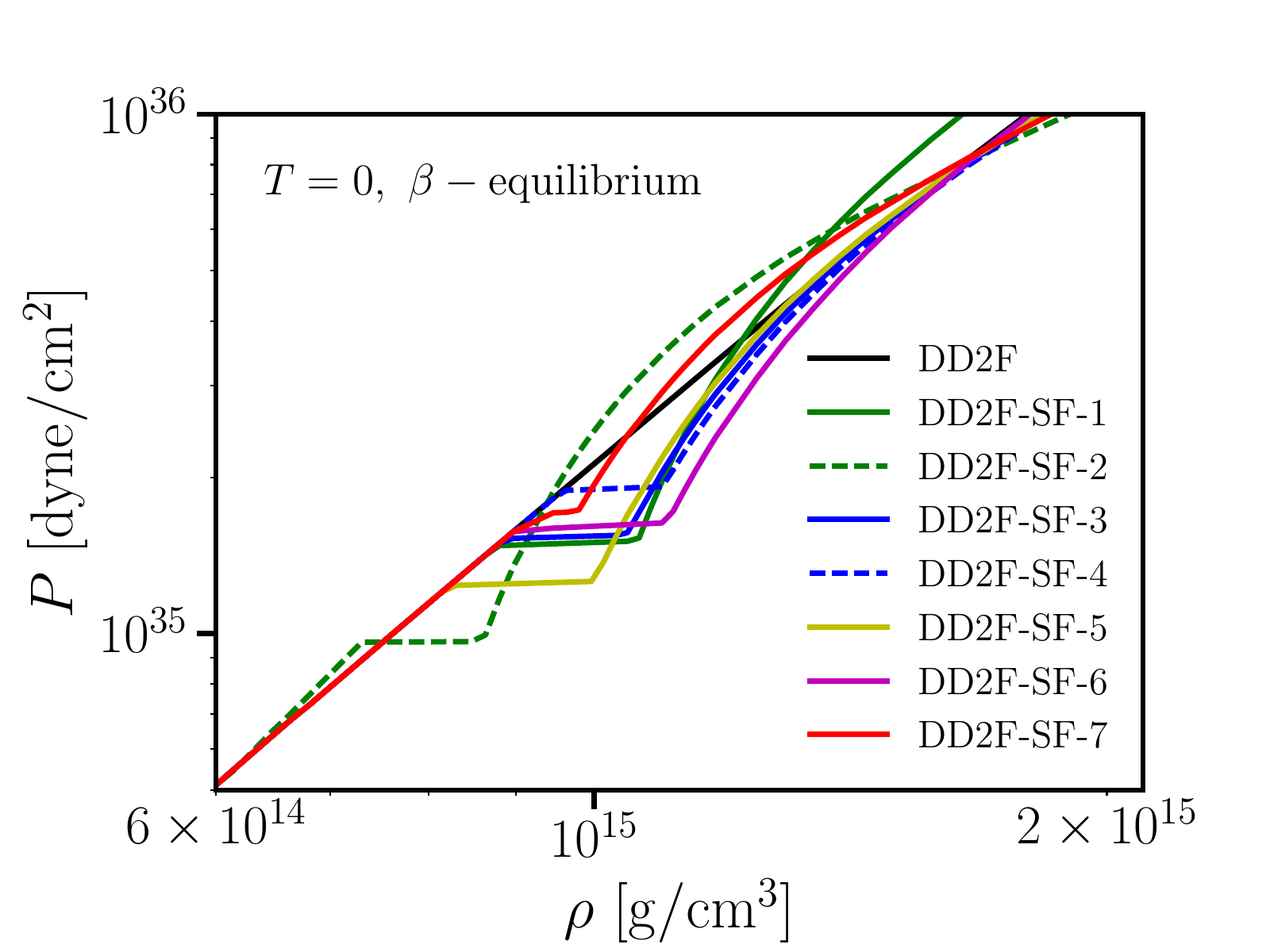}
  \caption{\label{fig:eos}Pressure as function of the rest-mass density for different EoSs discussed in this paper. The black curve is based on a purely hadronic relativistic mean-field model, while the other curves show EoSs which feature a first-order phase transition to quark matter. Figure taken from the Supplemental Material of~\cite{Bauswein2019}.}
\end{figure} 
\begin{figure}[h]
  \includegraphics[width=250pt]{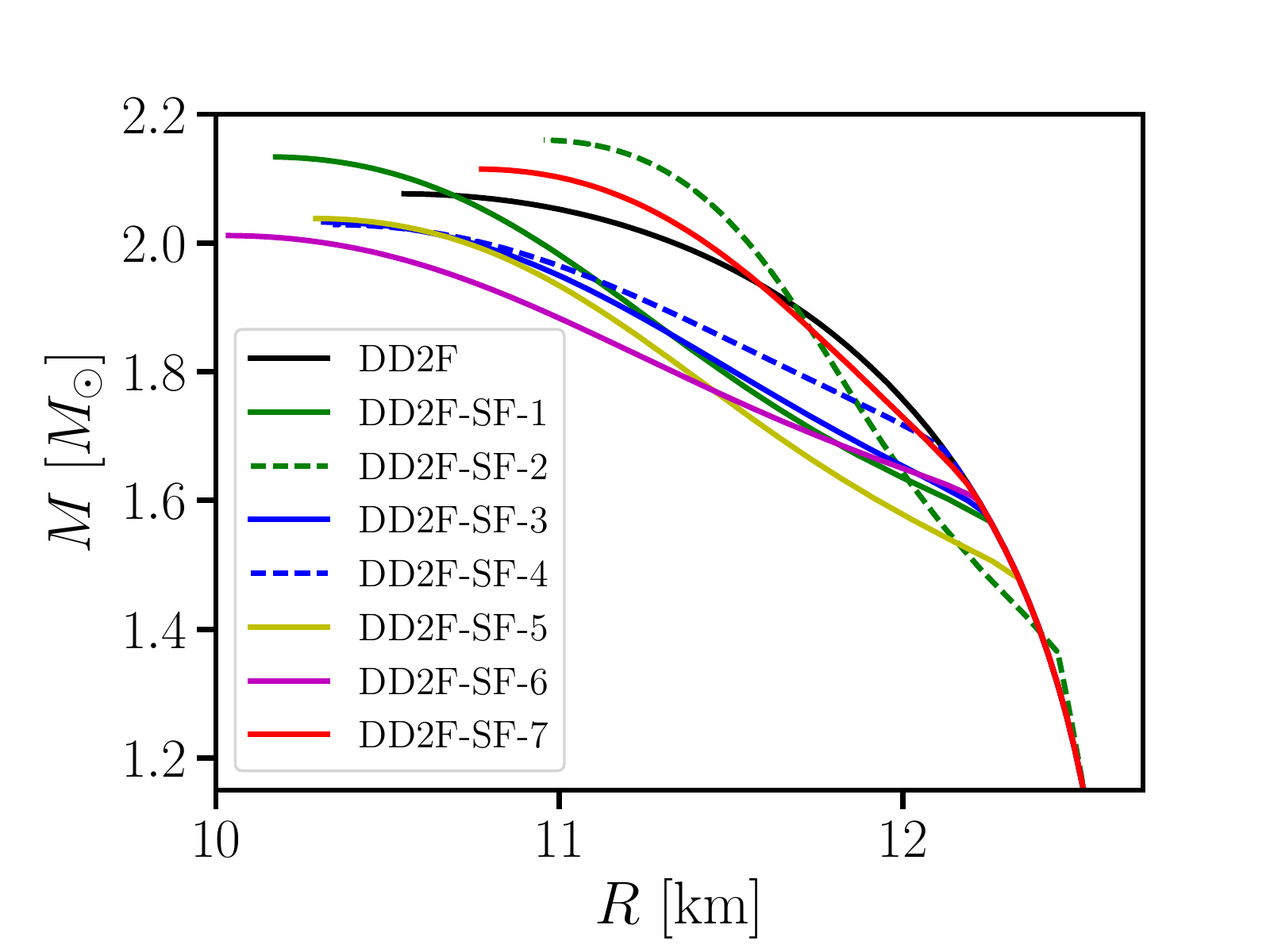}
  \caption{\label{fig:rm}Gravitational mass as function of the radius for different EoSs discussed in this paper. The lines correspond to the same EoSs as in Figure~\ref{fig:eos}. A first-order phase transition leads to a kink in the mass-radius relation.  Figure taken from the Supplemental Material of~\cite{Bauswein2019}.}
\end{figure}

Therefore, in this work we consider the full merger process including the postmerger phase to extract signatures of a phase transition to quark matter. On one hand we consider a specific nucleonic reference model DD2F describing the EoS at all densities. This EoS is described within a relativistic mean-field model with density-dependent coupling constants~\cite{Typel2005,Typel2010,Alvarez-Castillo2016}. This nucleonic EoS fulfills current constraints from nuclear physics and astrophysics, e.g.~\cite{Antoniadis2013,Danielewicz2002,Tsang2009,Tsang2018,Zhang2018,Krueger2013,Lattimer2013,Oertel2017,Abbott2017,Bauswein2017,Margalit2017,De2018,Abbott2018,Zhu2018a}. On the other hand we employ the same nucleonic model only at densities below a phase transition to deconfined quark matter. The quark matter is described by a phenomenological two-flavor string-flip model~\cite{Kaltenborn2017,Fischer2018,Bastian2018}, including scalar and vector interactions on the meanfield level. The phase transition from nucleonic matter to quark matter is obtained with a Maxwell construction, while considering chemical phase-equilibrium for baryon number $\mu_\mathrm B^\mathrm H = \mu_\mathrm B^\mathrm Q$ and charge number $\mu_\mathrm C^\mathrm H = \mu_\mathrm C^\mathrm Q$ independently. The resulting hybrid model leads to NS properties which are compatible with current constraints on NS radii and the maximum mass of nonrotating NSs.

By changing the parameters of the string-flip model, we obtain in total seven different realizations of hybrid EoSs based on the same nucleonic model (see~\cite{Fischer2018} for details on the models and~\cite{Bauswein2019} for the chosen parameters). These different EoSs are shown in Figure~\ref{fig:eos} together with the purely nucleonic reference model (black curve). We call these models DD2F-SF-n with $n=\{1,2,3,4,5,6,7\}$, where $n$ specifies a particular choice of string-flip parameters. The different models cover variations of the onset density of the phase transition between 0.466~$\mathrm{1/fm^3}$ and 0.580~$\mathrm{1/fm^3}$. Similarly, the density jump across the phase transition and the stiffness of the quark matter phase is varied within our sample of hybrid models. These variations are also apparent in the resulting mass-radius relations of nonrotating NSs displayed in Figure~\ref{fig:rm}.

The occurrence of the phase transition to quark matter at higher densities is clearly visible as a kink in the different mass-radius relations. These transitions take place at different masses depending on the onset density of the phase transition. For instance, for DD2F-SF-2 with the lowest onset density a quark core is present in NSs with masses above 1.37~$M_\odot$, whereas for DD2-SF-4 (with the highest onset density) quark matter appears only in nonrotating stars more massive than 1.68~$M_\odot$. Generally the occurrence of quark matter is connected with a compactification of the NSs, i.e. with smaller radii compared to the nucleonic reference model. This is understandable considering the density jump across the phase transition, which effectively resembles a strong softening of the EoS. However, for some of our models quark matter stiffens at higher densities such that some high-mass hybrid stars can have radii larger than those of the nucleonic reference model at the same mass. The mass-radius relations shown in Figure~\ref{fig:rm} are computed for cold, nonrotating NSs in neutrinoless beta-equilibrium.

We stress that our EoS models consistently describe the temperature and the composition dependence of nucleonic and quark matter. This is important because during merging temperatures of several 10 MeV can be reached. At such temperatures the phase boundaries are shifted compared to zero temperature, which may affect the merger dynamics.

In addition to these models we consider several other purely hadronic EoSs, which are based on different theoretical prescriptions of high-density matter. See~\cite{Bauswein2019} for more details. Employing a larger sample of purely hadronic models is important to demonstrate that a particular signature is unambiguously related to the occurrence of a first-order phase transition. An observed difference between a hadronic reference model and a hybrid model is not an unambiguous signature if a similar difference could result from another viable hadronic EoS.  
\subsection{Simulations and dynamics}

We focus on calculations of binary mergers of two stars with 1.35~$M_\odot$ starting from quasi-circular equilibrium orbits. The stars are initially at zero temperature and the composition is chosen such that matter is initially in neutrinoless beta-equilibrium. More details on the setup and the simulation tool can be found in~\cite{Bauswein2019} and references therein.
\begin{figure*}[h!]
  \includegraphics[width=250pt]{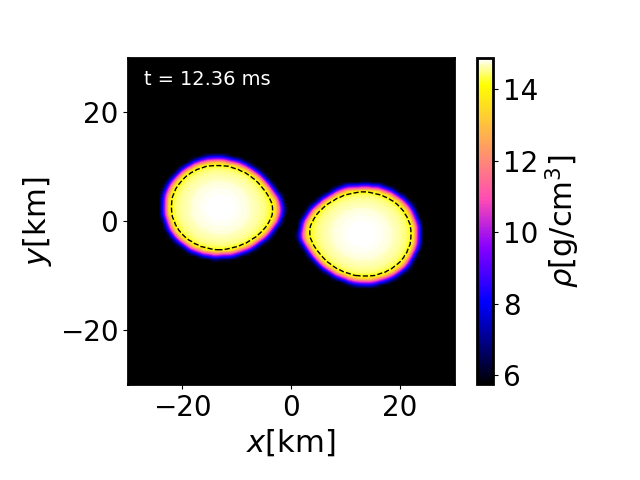}\includegraphics[width=250pt]{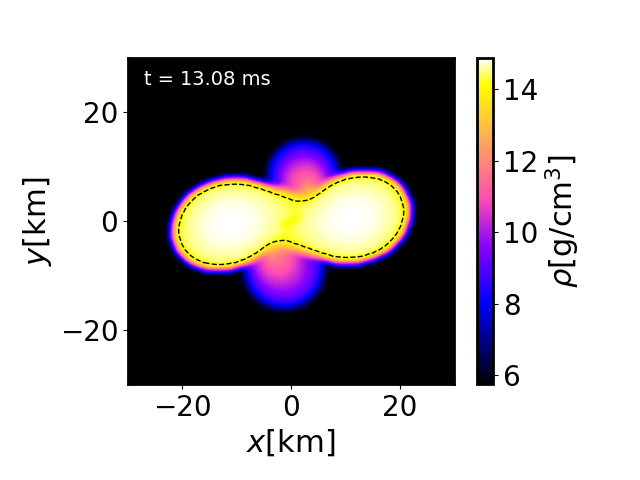}\\
  \includegraphics[width=250pt]{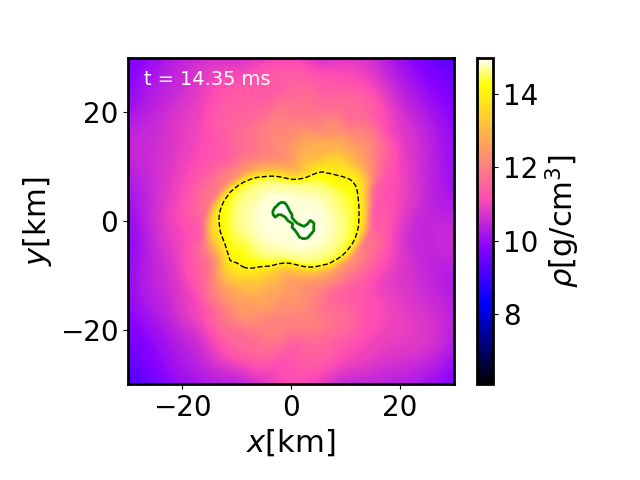}\includegraphics[width=250pt]{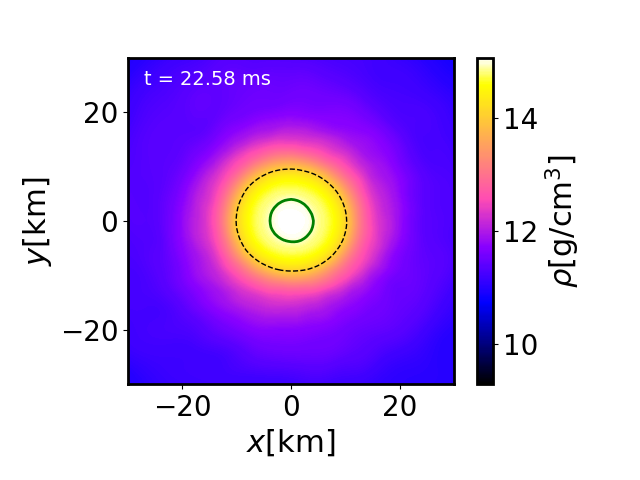}
  \caption{\label{fig:snap} Rest-mass density color-coded in the equatorial plane for a merger of two stars of 1.35~$M_\odot$ described by the DD2F-SF-1 EoS. The dashed line displays the density contour with $\rho=10^{14}~\mathrm{g/cm^3}$. The solid curve shows the density contour with $8.7\times10^{14}\mathrm{g/cm^3}$, which corresponds roughly to the onset density of the phase transition. Note that the density contour is only an approximate indicator of the phase transition because the onset density depends on the temperature. The time labels are shifted by 6.5~ms compared to the time axis in Figure~\ref{fig:rhomax}.}
\end{figure*}

Figure~\ref{fig:snap} shows the evolution of the rest-mass density $\rho$ in the equatorial plane for the simulation with DD2F-SF-1. The dashed line exhibits the density contour with $\rho=10^{14}\mathrm{g/cm^3}$. The green line indicates the area with a density of $8.7\times10^{14}\mathrm{g/cm^3}$, which corresponds to the onset density of DD2F-SF-1 at zero temperature and beta-equilibrium. The contour approximately illustrates the quark matter core bearing in mind that higher temperatures lead to lower onset densities of the phase transition. 
\begin{figure}[h]
  \includegraphics[width=250pt]{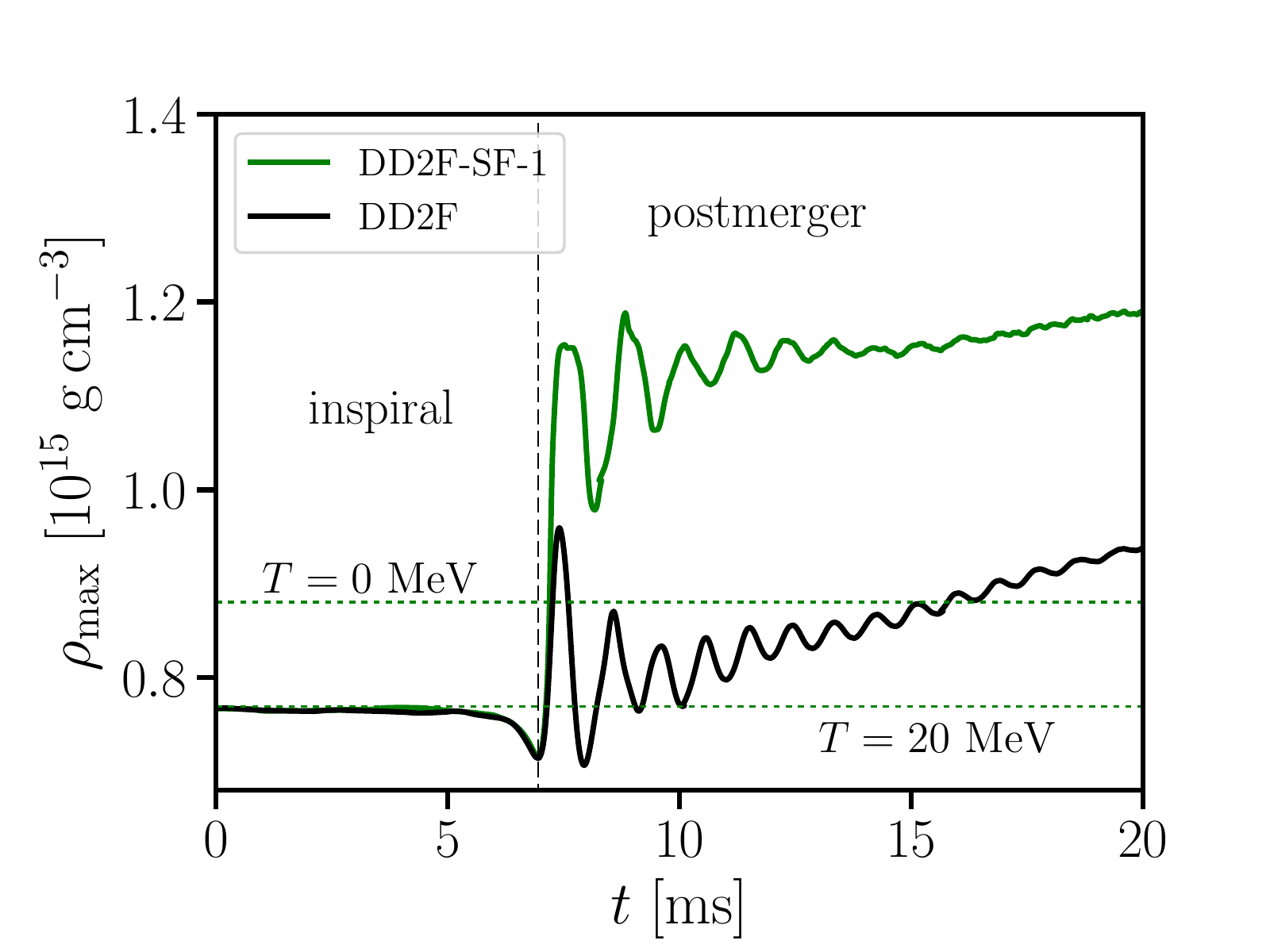}
  \caption{\label{fig:rhomax}Maximum rest-mass density as function of time in merger simulations of 1.35-1.35~$M_\odot$ binaries described by the purely hadronic EoS DD2F (black curve) and the hybrid EoS DD2F-SF-1 with a phase transition to quark matter (green curve). The onset density of the phase transition is indicated by a dotted horizontal line for $T=0$~MeV and for $T=20$~MeV. The horizontal dashed line indicates the time of merging.  Figure taken from~\cite{Bauswein2019}.}
\end{figure}

\begin{figure}[h]
  \includegraphics[width=250pt]{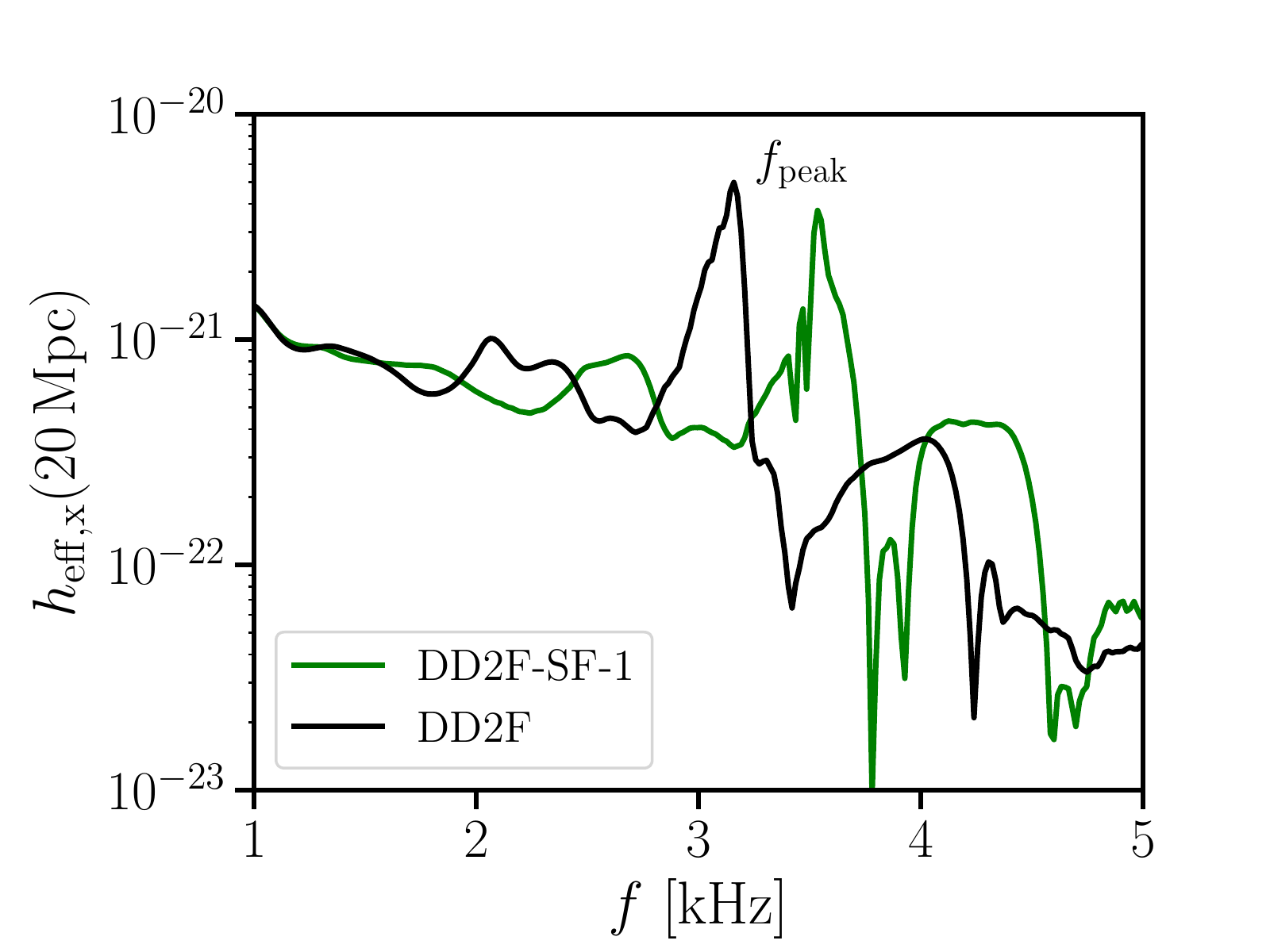}
  \caption{\label{fig:spectrum}GW spectrum for the two 1.35-1.35~$M_\odot$ merger simulations shown in Figure~\ref{fig:rhomax} at a distance of 20~Mpc. The black line corresponds to the GW signal extracted from the calculation with the purely hadronic EoS DD2F. The green curve displays the GW spectrum for the DD2F-SF-1 model, which features a first-order phase transition. Quark matter occurs directly after merging. $f_\mathrm{peak}$ labels the dominant postmerger oscillation frequency. Figure taken from~\cite{Bauswein2019}.}
\end{figure}

The upper left panel displays the two stars during the very last stage of the inspiral phase. Tidal deformations of the NSs are visible by eye. The densities are still below the onset density of the phase transition and no quark matter is present in this phase. Roughly half a millisecond later the two stars merge (upper right panel). During merging the densities and the temperatures in the merger remnant strongly increase and a quark matter core is formed in the center of the remnant. The central object is initially strongly oscillating and rapidly rotating. The rapid rotation is the main reason for the stability of the remnant, whose total mass exceeds the maximum mass of nonrotating NSs for the employed EoS model. On somewhat longer time scales of several tens of milliseconds the merger remnant settles into a more axi-symmetric configuration (lower right panel). In this stage the oscillations of the remnant are less pronounced and a sizeable quark matter core resides in the center of the remnant.

\begin{figure}[h]
  \includegraphics[width=250pt]{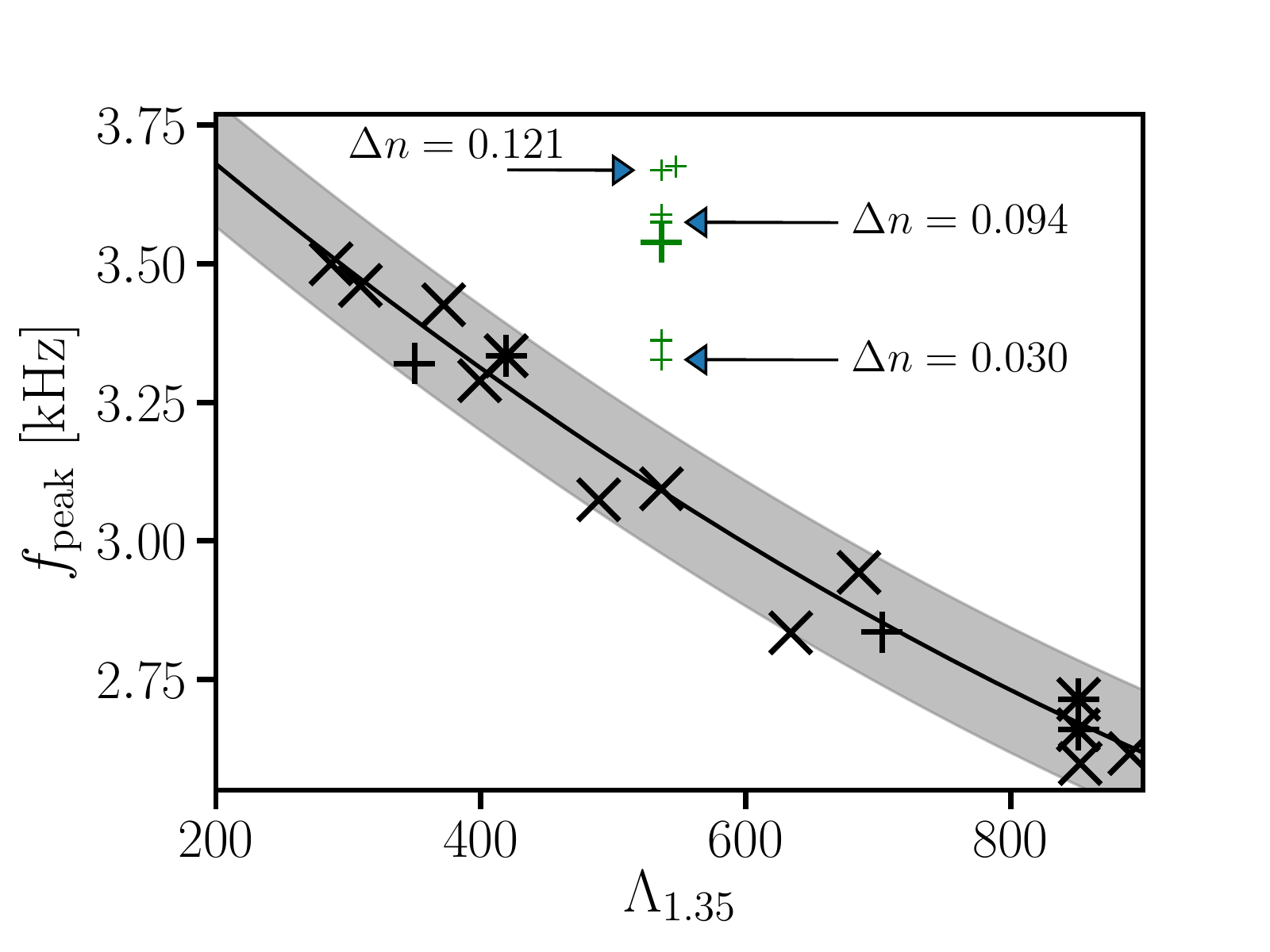}
  \caption{Postmerger GW frequency as function of the tidal deformability for 1.35-1.35~$M_\odot$ mergers described by different model EoSs. Black symbols display results for purely hadronic EoSs (except for the black plus signs which correspond to hybrid models of~\cite{Alford2005}, see main text). Green markers show the peak frequency for EoS models with a first-order phase transition to quark matter. The solid curve is a fit to the data excluding hybrid models and the gray band indicates the maximum deviation from the fit among the purely hadronic models. Figure taken from~\cite{Bauswein2019}.}
\label{fig:fpeak}
\end{figure}

\begin{figure}[h]
  \includegraphics[width=250pt]{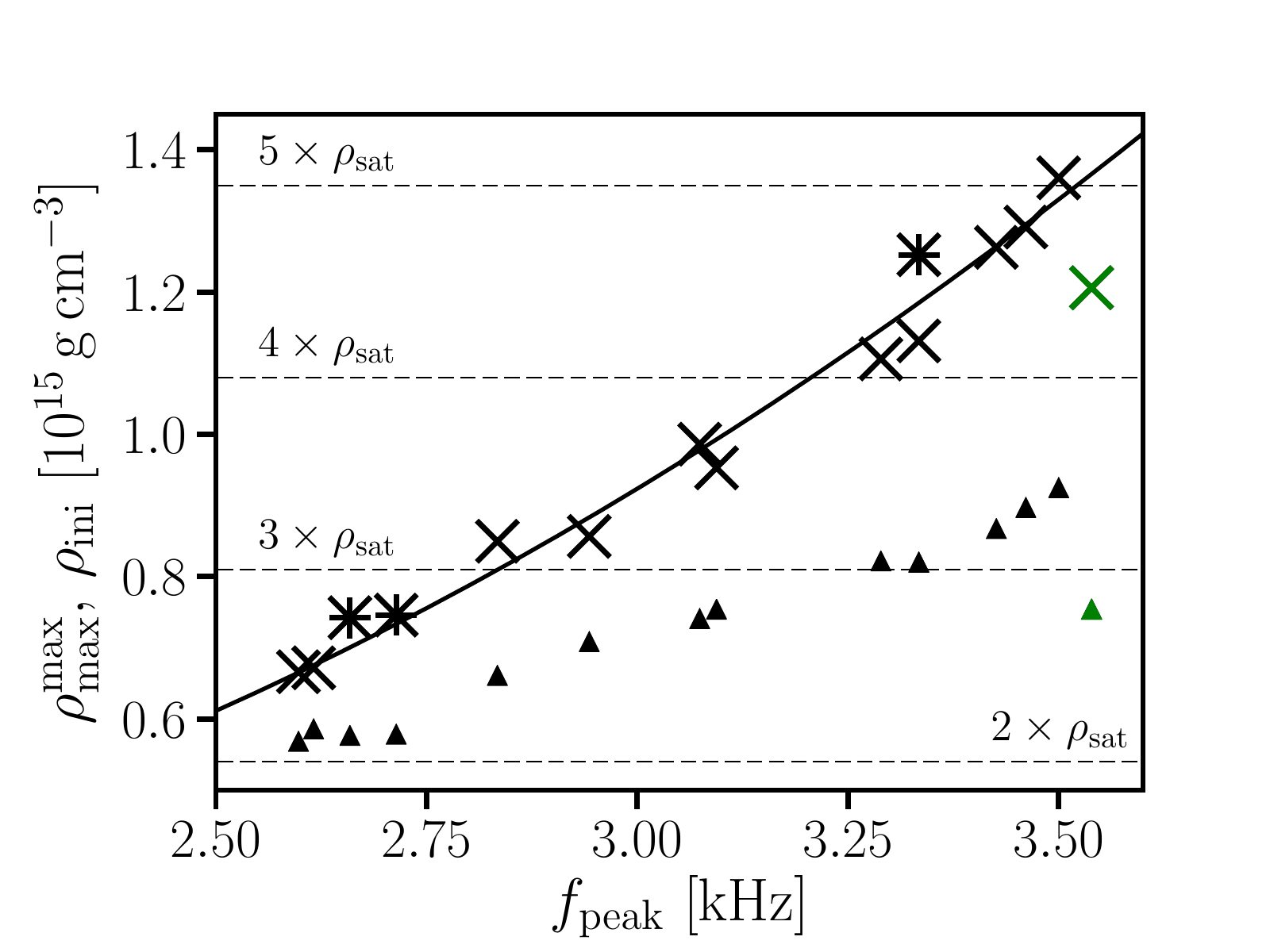}
  \caption{\label{fig:rho} Maximum rest-mass density during the early postmerger evolution of 1.35-1.35~$M_\odot$ mergers for different EoSs as function of the dominant postmerger GW frequency. Symbols have the same meaning as in Figure~\ref{fig:fpeak}. Solid curve is a fit to the data excluding models which undergo a first-order phase transition. The triangles indicate the initial central density in the progenitor stars. The figure demonstrates the strong density increase during merging, which is greater for softer EoSs corresponding to higher postmerger frequencies. Figure following Figure~4 in~\cite{Bauswein2019}.}
\end{figure}

The evolution of the remnant is also reflected in Figure~\ref{fig:rhomax} showing the maximum rest-mass density as function of time for the same simulation of DD2F-SF-1 (time shifted by about 6.5~ms) and for the purely nucleonic reference model DD2F. The two stars merge at about 7~ms (dashed vertical line). The inspiral phase proceeds virtually identically for both models since the densities remain below the onset density of the phase transition. Tiny deviations during the inspiral are due to statistical fluctuation of the initial data and are dynamically irrelevant. When the stars collide, the maximum density increases strongly and exceeds the onset density of the phase transition (for the hybrid EoS DD2F-SF-1). The occurrence of a phase transition to quark matter softens the EoS, which only stiffens at higher densities (see Figure~\ref{fig:eos}). The maximum density of DD2F-SF-1 is thus higher compared to the purely nucleonic model DD2F. The variations of $\rho_\mathrm{max}$ during the postmerger phase are connected to the quasi-radial oscillation of the remnant, see the discussion in~\cite{Bauswein2015,Bauswein2019a}. Following the classification scheme in~\cite{Bauswein2015} the evolution of DD2F would be identified as Type I.

Interestingly, the evolution of $\rho_\mathrm{max}$ for DD2F-SF-1 looks qualitatively somewhat different compared to other simulations of hadronic EoSs in the literature (see e.g. figures in~\cite{Hotokezaka2013a} and description in~\cite{Bauswein2015} for $\rho_\mathrm{max}$  and the minimum of the lapse function). Typically the collision and the first bounce of the original NS cores lead to oscillations of $\rho_\mathrm{max}$ such that in the first decompression phase the maximum density decreases nearly to the maximum density before merging (as for DD2F at $t=8$~ms, see also Figure 3 in~\cite{Hotokezaka2013a}). The amplitude of the oscillations is then damped and $\rho_\mathrm{max}$ grows on average. In the case of DD2F-SF-1 the first decompression phase does not reduce $\rho_\mathrm{max}$ as strong. Instead the amplitude of the oscillations is weaker while the average of the oscillations in $\rho_\mathrm{max}$ is higher, which likely reflects the presence of the quark matter core. This behavior is also different from the other two types of postmerger dynamics (Type II and Type III) as defined in~\cite{Bauswein2015}, where the quasi-radial oscillations are less pronounced and instead a low frequency modulation of $\rho_\mathrm{max}$ is in addition present. It remains to be seen if the evolution of the hybrid model in Figure~\ref{fig:rhomax} is a generic feature of all models with strong first-order phase transitions to quark matter. 

\subsection{Gravitational-wave emission}

The different evolution of the two systems (DD2F and DD2F-SF-1) is also visible in the GW signal. Figure~\ref{fig:spectrum} shows the GW spectrum. Up to a frequency of about 1.7~kHz the spectra are very similar because the low frequency part of the spectra is strongly dominated by the inspiral phase, which proceeds identically in both simulations. The high-frequency regime of the spectra however is shaped by the postmerger evolution, which exhibits strong differences. In particular, the dominant oscillation frequency $f_\mathrm{peak}$ of the remnant, which is visible as a pronounced peak in the kHz range, is shifted to higher frequencies for DD2F-SF-1 compared to the nucleonic reference model DD2F. This behavior is understandable since the phase transition to quark matter leads to a more compact remnant, which thus oscillates at higher frequencies. 

A frequency shift of the dominant postmerger oscillation on its own may not yet be characteristic for a phase transition to quark matter. In principle, a hadronic EoS which is significantly softer than our reference model DD2F could lead to similarly high frequencies in the postmerger phase for the same total binary mass (see e.g.~\cite{Bauswein2012a}). However, the frequency shift is characteristic if one compares it to the tidal deformability of the inspiralling NSs.

The tidal deformability $\Lambda$ is the NS property which describes finite size effects in the premerger phase, i.e. the accelerated inspiral of larger NSs compared to less extended bodies or even black holes. As the tidal deformability affects the phase evolution of the binary, it can be measured from the inspiral GW signal~\cite{Hinderer2008,Read2013,Abbott2017,TheLIGOScientificCollaboration2018a,Abbott2018,De2018}. 

Figure~\ref{fig:fpeak} shows the comparison between the tidal deformability and the dominant postmerger GW frequency $f_\mathrm{peak}$ for binary systems of two stars with 1.35~$M_\odot$. All purely hadronic models of our large, representative sample follow a tight relation between $\Lambda$ and $f_\mathrm{peak}$. The maximum deviation from this relation is only about 100~Hz for EoS models without strong first-order phase transition. In contrast to these purely hadronic EoSs, those models which undergo a strong phase transition to deconfined quark matter (green symbols), result in a much stronger deviation from the tight $\Lambda$-$f_\mathrm{peak}$ relation of up to $\sim0.5$~kHz. 

These deviations are sufficiently strong that they will be detectable in future events. Calculations with simulated injections have shown that $f_\mathrm{peak}$ can be measured with a precision of a few tens of Hz for near-by events if the detector sensitivity is increased in the next years. For instance, for binary systems at distances similar to GW170817 currently projected upgrades to the existing detectors suffice~\cite{Clark2014,Clark2016,Chatziioannou2017,Yang2018,Torres-Rivas2019,Martynov2019}. Also, the tidal deformability will be measured with better accuracy compared to GW170817. 

We argue that identifying deviations in the $\Lambda$-$f_\mathrm{peak}$ relation would provide very strong evidence for a first-order phase transition taking place at a few times nuclear density. Importantly, we show by the consideration of a large set of purely hadronic EoSs including models of hyperonic matter, that such deviations cannot result from models without strong phase transition. Two of our models employing piecewise polytropes~\cite{Read2009a} mimic the transition to quark matter through a more continuous transition~\cite{Alford2005} (black crosses). Those calculations do not lead to significant deviations in the  $\Lambda$-$f_\mathrm{peak}$ correlation. Also the calculations with models that include hyperonic matter~\cite{Banik2014,Fortin2018,Marques2017} do not show characteristic differences compared to purely nucleonic EoSs. We thus argue that only a sufficiently strong phase transition with a strong density jump can yield such a clear, unambiguous signature of a phase transition in the GW signal of NS mergers. In our study we focus on first-order phase transitions with strong density jumps. Obviously, any kind of transition that effectively mimics such a type of transition, although it may formally not be first-order, may result in a similar signature.

Finally, we remark that a detection of the dominant postmerger frequency $f_\mathrm{peak}$ will also allow to constrain the density at which a phase transition does occur or not. Figure~\ref{fig:rho} relates the peak frequency with the maximum rest-mass density $\rho_\mathrm{max}^\mathrm{max}$ which is found during the first milliseconds of the remnant evolution. $\rho_\mathrm{max}^\mathrm{max}$ is a quantity which is not directly observable, but we can extract it from the simulations. For instance, for DD2F we find a maximum density of about $\rho_\mathrm{max}^\mathrm{max}=9.5\times10^{14}~\mathrm{g/cm^3}$ at about 7.5~ms. We focus here on the maximum of the early density evolution. Possibly, the density in the remnant could exceed this first maximum during the later evolution. However, at later times the postmerger GW emission is already damped and may not be strongly influenced by the evolution in this stage. 

By defining  $\rho_\mathrm{max}^\mathrm{max}$ as maximum during the early evolution, it represents the density regime that affects the GW postmerger emission. In turn, a measured peak frequency reveals the density regime that is probed by the remnant. If in a future event a peak frequency is measured which is consistent with the $\Lambda$-$f_\mathrm{peak}$ relation, this would exclude a strong phase transition in the merger remnant. Employing the relation in Figure~\ref{fig:rho}, it is possible to determine up to which density such a transition did not occur. Obviously the absence of a  signature of a phase transition as we identified in~\cite{Bauswein2019}, does not allow any conclusion about the presence and nature of a phase transition at higher densities which are not encountered in the remnant. However, Figure~\ref{fig:rho} does allow to quantify up to which density a strong phase transition can be excluded. This will be a very important information for nuclear physics, since the relation in Figure~\ref{fig:rho} determines up to which density matter can be described by nuclear physics methods as opposed to a modeling of deconfined quark matter.

If a signature of a phase transition is found, Figure~\ref{fig:rho} roughly determines the density below which the transition took place. Figure~\ref{fig:rho} also provides the central densities of the progenitor stars (triangles). The figure demonstrates that the densities in the merger remnant are generally higher than in the inspiralling stars. Hence, the merger remnant and its GW emission probe a different regime of the high-density EoS. The postmerger evolution and gravitational radiation are particularly interesting to investigate the EoS at very high densities. This quantifies the complementary information of inspiral and postmerger stages and stresses that both phases are important to learn about the properties of high-density matter. Moreover, these data reveal that the density increase from the inspiral to the merger phase is stronger for softer EoSs considering fixed binary masses. (The peak frequency is higher for softer EoSs.) 

Both quantities, the tidal deformability and the dominant postmerger frequency, have been shown to be measurable with sufficient precision in future events with high signal-to-noise ratio~\cite{Clark2014,Clark2016,Chatziioannou2017,Torres-Rivas2019}. Other signal characteristics like the exact phase evolution, the remnant life time or drifts of frequencies are much harder if not impossible to derive even with upgraded detectors. Also, those quantities may be stronger affected by the numerical errors due to the limited resolution and by incomplete physics in the simulations. Finally, it is not clear whether a phase transition leaves an unambiguous signature in these quantities or whether for instance a particularly soft hadronic EoS could generate very similar features. As we argued above this has to be shown by means of a large representative sample of hadronic candidate EoSs.

\section{Mass ejection in NS mergers undergoing phase transitions}

\begin{figure}[h]
  \includegraphics[width=250pt]{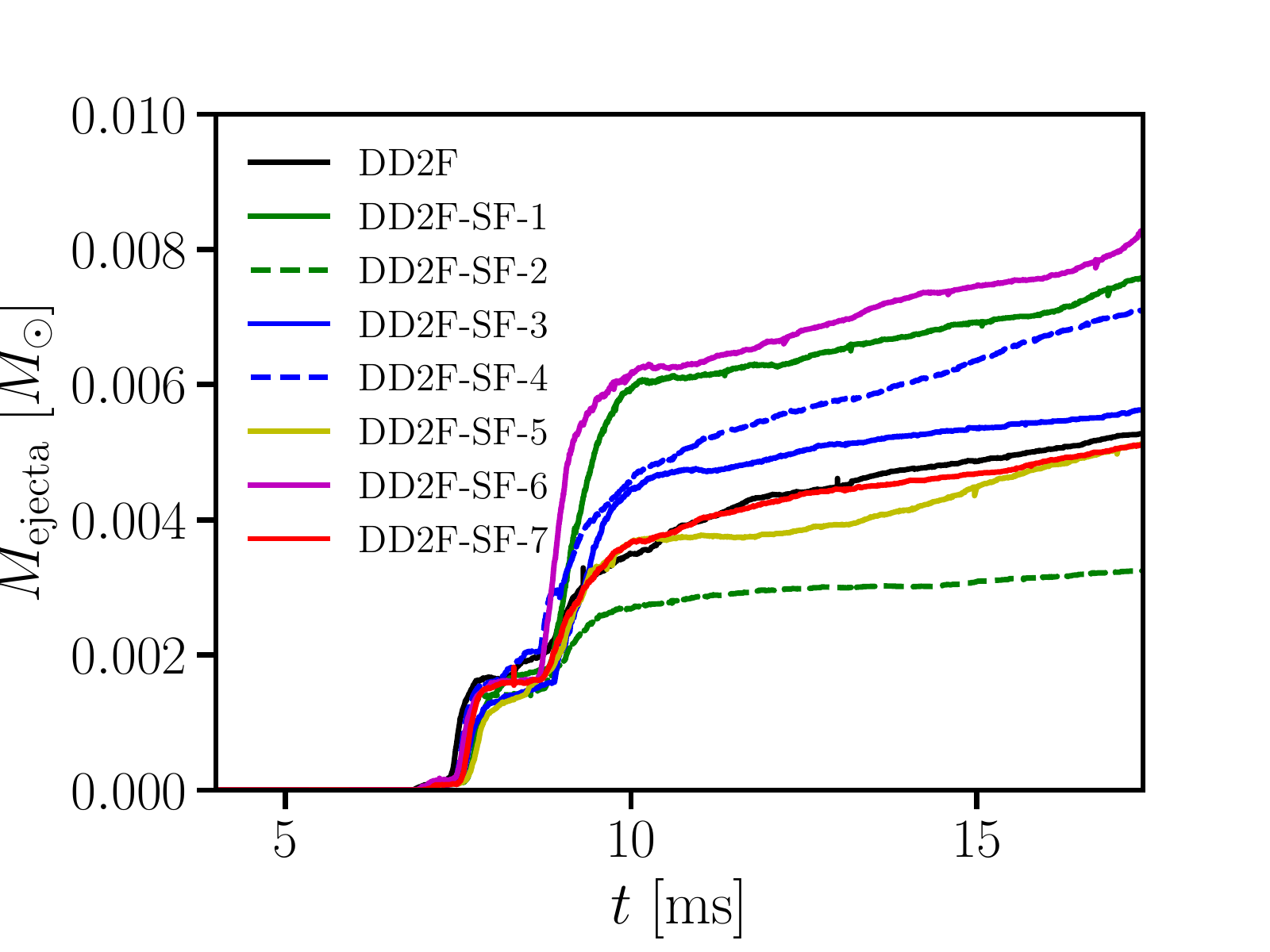}
  \caption{\label{fig:ejecta} Time evolution of the dynamical ejecta mass for 1.35-1.35~$M_\odot$ mergers. The line colors and styles have the same meaning as in Figure~\ref{fig:eos}. The black curve displays the evolution for the purely hadronic EoS DD2F, whereas all other models undergo a first-order phase transition during merging. The stars collide at about 7~ms.}
\end{figure}

We briefly discuss the effect of first-order phase transitions on the mass ejection of NS mergers, which is relevant for the formation of heavy elements through the r-process and for the electromagnetic counterpart which is powered by radioactive decays. We focus here on 1.35-1.35~$M_\odot$ binaries, which have a total mass roughly comparable to GW170817. Figure~\ref{fig:ejecta} shows the amount of unbound material for 1.35-1.35~$M_\odot$ mergers in the first 10~ms after merging (the stars merge at about 7~ms). The purely hadronic reference model DD2F leads to about 0.005~$M_\odot$ of dynamical ejecta (i.e. not including matter which becomes unbound on longer time scales from a torus around the central object). Models with a first-order phase transition, which lead to the appearance of a quark-matter core after merging, do not show strong differences compared to the purely hadronic reference EoS (black curve). A scatter of a few $10^{-3}~M_\odot$ between different EoSs may not be characteristic for hybrid models, as a similar variation is found between different purely hadronic EoSs which lead to similar NS radii (compare to the upper left panel in Figure~3 of~\cite{Bauswein2013a}).

We conclude that the occurrence of quark matter may not too strongly affect the properties of an electromagnetic counterpart and element formation. Also other characteristic ejecta properties like the outflow velocity in hybrid models do not exhibit any peculiar behavior compared to purely hadronic models. The differences are rather quantitative than qualitative. More work is required to understand whether or not a phase transition leads to characteristic differences. Possibly a phase transition may have a more subtle impact on the ejecta for instance through altered neutrino emission, which modifies the electron fraction in the ejecta. This may influence details of the r-process and may also have some impact on the electromagnetic signal. With the current set of simulations we cannot further investigate this aspect as neutrino interactions were neglected. The consideration of neutrinos would require detailed knowledge about the weak interaction rates of quark matter.  We expect only quantitative changes as the quark matter core resides deep inside the center of the merger remnant, whereas the neutrinosphere lies further outside and is shielded by nuclear matter from the central region. But it remains to be seen whether the stronger compactification of the remnant due to the phase transition leads to a characteristic change of the neutrino emission. Also, it should be explored in more detail whether hybrid models result in significant differences in the long-term evolution in comparison to purely hadronic models.

Based on our calculations and this relatively simple analysis, we conclude that the electromagnetic counterpart of GW170817 is compatible with the occurrence of quark matter. The similarity between purely hadronic mergers and systems which undergo a phase transition, also implies that it may be difficult to infer the presence of a phase transition from the bulk properties of the electromagnetic transient. This additionally stresses the importance of detecting postmerger GW emission to learn about the occurrence of a phase transition in NSs as described above.

\section{Summary and discussion}

Based on a minimum set of assumptions we showed that the interpretation of the electromagnetic emission of GW170817 and its measured total binary mass imply a robust lower limit on NS radii of about 10.7~km. This constraint is a result of the fact that NSs with smaller radii would induce the direct collapse to a black hole immediately after merging. This would very likely be associated with reduced mass ejection, which is incompatible with the observation that indicates a rather high ejecta mass. The underlying assumptions of this constraint can be verified in future observations as our understanding of the electromagnetic emission further grows. Higher binary masses in future events with a similarly bright optical display would further strengthen our limit. We emphasize that the currently available data does not allow for stronger constraint in contrast to some other claims in the literature. EoSs with radii larger than about 10.7~km are compatible with the current observations.

Moreover, a future NS merger with a much dimmer electromagnetic counterpart would indicate a prompt collapse of the merger remnant. In that case the measured total binary will yield an upper limit on NS radii and importantly also a robust upper limit on the maximum mass of nonrotating NSs. We stress that our new method for EoS constraints from NS mergers can even employ detections with low signal-to-noise ratio if those events provide an indication of the merger outcome by the electromagnetic emission. The procedure is thus very promising for further EoS constraints in the near future.

Beyond currently possible constraints and near-future prospects, we also describe how the presence of a phase transition alters the GW signal of NS mergers and may be detectable in several years from now when the available GW instruments reach their design sensitivity or receive an upgrade. We provide strong evidence that the feature which we identified as being characteristic for the occurrence of quark matter, can only be caused by the presence of a sufficiently strong phase transition. Specifically, we show that the dominant postmerger GW frequency is shifted to higher frequencies if a phase transition takes place. 

Finally, we compare the mass ejection in purely hadronic NS mergers and systems which undergo a phase transition to quark matter. We do not identify any qualitative differences which are potentially indicative for a phase transition considering, however, only bulk properties like the total ejecta mass. We argue that, based on these simple considerations, GW170817 and the associated electromagnetic emission are in principle compatible with EoSs which feature a strong phase transition. 

\section{ACKNOWLEDGMENTS}

AB acknowledges support by the European Research Council (ERC) under the European Union's Horizon 2020 research and innovation programme under grant agreement No. 759253 and the German Research Foundation (DFG) via the Collaborative Research Center SFB 881 ``The Milky Way System''. NUFB and TF acknowledge support from the Polish National Science Center (NCN) under grant no. UMO-2016/23/B/ST2/00720. DB acknowledges support through the Russian Science Foundation under project No. 17-12-01427 and the MEPhI Academic Excellence Project under contract No.~02.a03.21.0005. We acknowledge stimulating discussions during the EMMI Rapid Reaction Task Force: The physics of NS mergers at GSI/FAIR and the support of networking activities by the COST Actions CA15213 ``THOR'', CA16117 ``ChETEC'', CA16214 ``PHAROS'' and CA16104G ``GWVerse''. OJ is supported by the Special Postdoctoral Researchers (SPDR) program and the iTHEMS cluster at RIKEN. NS is supported by the ARIS facility of GRNET in Athens (GWAVES and GRAVASYM allocations). HTJ is grateful for support by the German Research Foundation (DFG) through Collaborative Research Center SFB 1258 ``Neutrinos and Dark Matter in Astro- and Particle Physics'' (NDM) and the Excellence Cluster Universe (EXC 153; http://www.universe-cluster.de/). 
The Flatiron Institute is supported by the Simons Foundation.  JC acknowledges support from NSF award PHY-1505524.





\end{document}